\documentclass[conference]{IEEEtran}
\IEEEoverridecommandlockouts

\usepackage{cite}
\usepackage{amsmath,amssymb,amsfonts}
\usepackage{algorithmic,subcaption}
\usepackage{graphicx}
\usepackage{textcomp,url,multicol,multirow}
\usepackage{xcolor}
\usepackage{tikz}
\def\BibTeX{{\rm B\kern-.05em{\sc i\kern-.025em b}\kern-.08em
    T\kern-.1667em\lower.7ex\hbox{E}\kern-.125emX}}
\begin{document}

\title{Can Highlighting Help GitHub Maintainers Track Security Fixes?
}

\author{\IEEEauthorblockN{Xueqing Liu}
\IEEEauthorblockA{
\textit{Stevens Institute of Technology}\\
xliu127@stevens.edu}
\and
\IEEEauthorblockN{Yuchen Xiong}
\IEEEauthorblockA{\textit{Nanjing University} \\
yuchenxiong@smail.nju.edu.cn}
\and
\IEEEauthorblockN{Qiushi Liu}
\IEEEauthorblockA{\textit{Zhejiang University} \\
qiushi3@illinois.edu}
\and
\IEEEauthorblockN{Jiangrui Zheng}
\IEEEauthorblockA{\textit{Stevens Institute of Technology} \\
jzheng36@stevens.edu}
}

\maketitle

\begin{abstract}
In recent years, the rapid growth of security vulnerabilities poses great challenges to tracing and managing them. For example, it was reported that the NVD database experienced significant delays due to the shortage of maintainers~\cite{nvd_delay}. The delays in updating the patch link information can pose significant security risks, as organizations may remain vulnerable to known exploits before the patch link is updated. Furthermore, the delay creates challenges for third-party security personnel (e.g., administrators) to trace the information related to the CVE. 

To help security personnel trace a vulnerability patch, in this work, we build a retrieval system that automatically retrieves the patch in the repository. While existing work leverages deep learning models such as CodeBERT for tracing the vulnerability fix, one major challenge with deep learning models is the lack of transparency. Furthermore, since the retrieved code is usually very long, it can be challenging for the security maintainer to understand the model output. 

Inspired by existing work on explainable machine learning, we ask the following research question: can explanations help security maintainers make decisions in patch tracing? First, we investigate using LIME~\cite{ribeiro2016should} (a widely used explainable machine learning method) to highlight the rationale tokens in the commit message and code. However, we observe that LIME often selects non-informative words, especially for the code. Therefore we propose an explanation method called TfIdf-Highlight, which leverages the Tf-Idf statistics to select the most informative words in the repository and the dataset. 

We evaluate the effectiveness of highlighting using two experiments. First, we compare LIME and TfIdf-Highlight using a faithfulness score (i.e., sufficiency and comprehensiveness) defined for ranking~\cite{deyoung2019eraser}. Faithfulness is a standard metric for evaluating explainable machine learning. We find that TfIdf-Highlight significantly outperforms LIME's sufficiency scores by 15\% and slightly outperforms the comprehensiveness scores. Second, we conduct a blind human labeling experiment by asking the annotators to guess the patch under 3 settings (TfIdf-Highlight, LIME, and no highlight). We find that the helpfulness score for TfIdf-Highlight is higher than LIME while the labeling accuracies of LIME and TfIdf-Highlight are similar. Nevertheless, highlighting does not improve the accuracy over non-highlighting. 

\end{abstract}


\section{Introduction}

\begin{figure}[h]
 \centering
 \begin{subfigure}{0.45\textwidth}
        \centering
        \includegraphics[width=\linewidth]{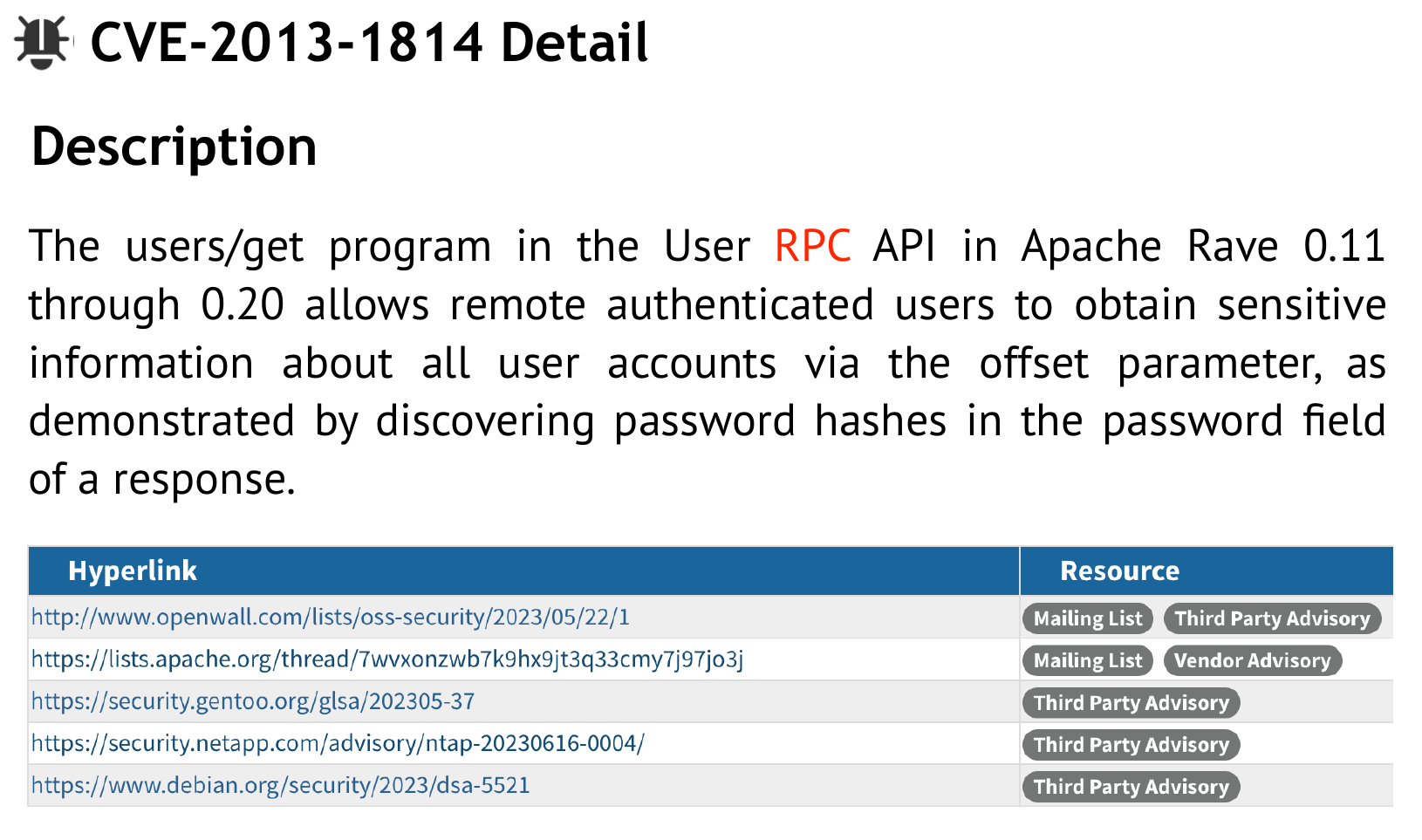}
       \caption{An example of patch missing (Nov 2024) in the NVD database}
    \end{subfigure}
    \vspace{0.21in}
    \\
 \begin{subfigure}{0.45\textwidth}
        \centering
        \includegraphics[width=\linewidth]{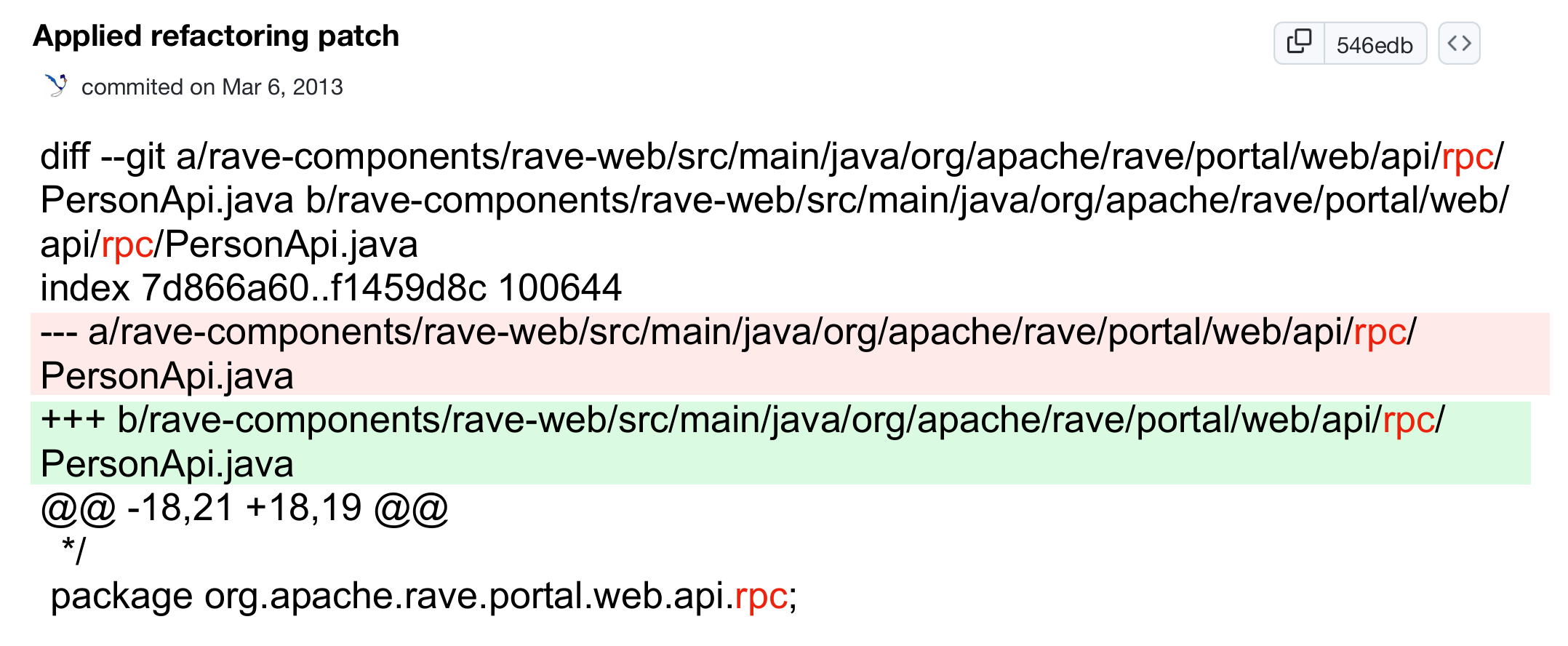}
       \caption{The relevant patch of CVE-2013-1814}
    \end{subfigure}
    \caption{An example of NVD's missing patch link\label{fig:cve_example}}
\end{figure}

In recent years, Open-Source Software (OSS) has been widely adopted, deployed, and utilized across industries. One of the biggest challenges that OSS faces is the prevalence of security vulnerabilities. For example, in 2017, Equifax suffered from a data breach, compromising the personal information of 147 million users, causing millions of dollars of loss, and exposing individuals to potential identity theft~\cite{equifax}. The attack was exploited because Equifax did not patch an Apache Strut vulnerability whose patch was already available. As a result, it is important for OSS users to be alerted of the updates of vulnerabilities and apply the patches in time. For the same reason, if a patch is available, it is important for security advisory databases to update the patch information in time. One of the most popular public vulnerability advisory databases for security vulnerabilities is the National Vulnerability Database (NVD)~\cite{nvd}. It is a common practice for users to rely on NVD for monitoring related information regarding the updates, patches, etc. However, in recent years, due to the rapid growth of CVE requests, NVD has experienced a drastic delay in processing the CVE submitters' requests~\cite{nvd_delay,nvd_delay2,nvd_delay3}. For example, Figure~\ref{fig:cve_example} shows the NVD page of CVE-2013-1814, where the patch link is still missing after 11 years. Nevertheless, this patch actually exists (created in March 2013) and has been identified by the GitHub Advisory database in 2023~\cite{CVE-2013-1814_github_pr}. The patch is shown in Figure~\ref{fig:cve_example}, and its relevance with the CVE can be justified by the overlapping token "\emph{rpc}". 

Besides posing security threats, another challenge caused by the delay of NVD in updating the patch is the information validation of CVE-related maintenance tasks. For example, the GitHub advisory database~\cite{githubAD} relies on users to submit pull requests for the updates of meta-data information about a CVE (e.g., affected package name, affected version), then GitHub maintainers review and approve these pull requests before they are merged in the database. However, by reviewing tens of pull requests, we observe that the pull requests submitted by users often lack the information source, causing challenges for maintainers in understanding these pull requests and delays in approving these pull requests. One example is the request for changing the affected package name in CVE-2013-1814~\cite{CVE-2013-1814_github_pr}. Since the requester did not submit a link to the original patch, the GitHub maintainer has to manually search the patch by herself to verify the request. While security personnel can manually search the web/repository for the patches, this approach is time-consuming and may be prone to bias.

In this work, we identify that a critical step for assisting security personnel (e.g., database maintainers, and submitters) in maintaining CVE-related information is to build a retrieval system for the CVE-related code, especially for retrieving the patch. The closest work to our problem is Sun et al.~\cite{sun2024tracing}, who build an automated retrieval system for tracing the relevant fix at the file level. Their method employs neural network models including CodeBERT~\cite{feng2020codebert} and RoBERTa~\cite{liu2019roberta}. The neural networks are used for retrieving the files most relevant to a CVE based on the CVE description as the query. However, one major challenge with the neural network approach is the lack of explainability~\cite{ribeiro2016should,pei2017deepxplore,guo2018lemna}. That is, while CodeBERT can output a ranked list of files, it cannot explain why it makes the decision of one ranked list and not another. On the other hand, since code commits typically contain a large number of tokens (the median number of tokens in our dataset in Table~\ref{tab:stat} is 1,840), understanding the decision of a CodeBERT model is even more challenging. 

To assist security personnel make informed decisions, this paper proposes to build an explainable model for retrieving the patch. Existing work on explainable machine learning (XML) has studied how to improve the informativeness of human decision-making when provided with the result of a model. While there are various ways to explain a machine-learning decision, one widely used technique is to select a few tokens that are the most important for the model's decision-making or the user's mental model. Previous work on explainable machine learning (XML)~\cite{ribeiro2016should} report that by highlighting important tokens in the input (i.e., the explanation), it is easier for human users to accurately annotate the class label. Motivated by previous work's findings, this paper studies the following question: can explainable machine learning help humans retrieve the relevant patch of a CVE? In Figure~\ref{fig:cve_example}, we show an envisioned interface for this purpose. While the system retrieves the relevant patch, it also highlights the matched token "\emph{rpc}". 

To answer the proposed question, our study includes the following steps. First, since no existing datasets can be directly used for the setting of our problem, we construct a new dataset. We reuse the (CVE, patch) pairs from existing datasets, then we leverage the version information (e.g., CPE) provided by the security database to reduce the search space to a smaller number of commits. Second, following Sun et al.~\cite{sun2024tracing}, we train a language model (e.g., CodeBERT) that retrieves commits (i.e., commit message and diff code) given the CVE description. Third, we implement two explainable machine-learning algorithms for the retrieval model. The first algorithm is LIME~\cite{ribeiro2016should}, a widely applied explainable machine learning algorithm. However, we observe that LIME sometimes selects non-informative words. Therefore we further introduce another explainable machine learning algorithm called TfIdf-Highlight. Fourth, we evaluate the effectiveness of explainable machine learning algorithms with the faithfulness score, a widely used metric for XML~\cite{deyoung2019eraser,jacovi2020towards} as well as human labeling experiments. 

The findings of our experiments are summarized as follows. First, by comparing the faithfulness scores (sufficiency and comprehensiveness) of LIME vs TfIdf-Highlight, we find that TfIdf-Highlight significantly lowers the sufficiency scores of LIME by around 15\% in almost all settings. That is, the highlighted words selected by Tf-Idf can better reflect the model's true decision-making process. Second, in the human labeling experiment, we ask three annotators to blindly select the patch among a group of commits. Each group either uses LIME, TfIdf-Highlight, or no highlight, and the groups are mixed so the annotator cannot tell between LIME and TfIdf-Highlight. The results of the human labeling experiments show two things. First, TfIdf-Highlight is more helpful than LIME Highlight. Second, highlighting does not result in a higher labeling accuracy. The reason is that while the highlighted tokens can help with eliminating token-level mismatches (e.g., file names, variable names), they cannot explain the semantic relatedness between the CVE description and the commit. 

In this paper, we make the following contributions:

\begin{itemize}
    \item We propose the first study on using an explainable retrieval system to assist vulnerability patch tracing;
    \item We collect a dataset for vulnerability patch tracing. The dataset will be made publicly available;
    \item We propose an explainable machine learning algorithm called TfIdf-Highlight that can highlight informative words in the CVE description and the patch. TfIdf-Highlight significantly outperforms LIME in the faithfulness score;
\end{itemize}









\section{An Empirical Study on the Availability of Patch Links in NVD}
\label{sec:empirical_study}


\begin{figure}[h]
 \centering
    \includegraphics[width=0.4\textwidth]{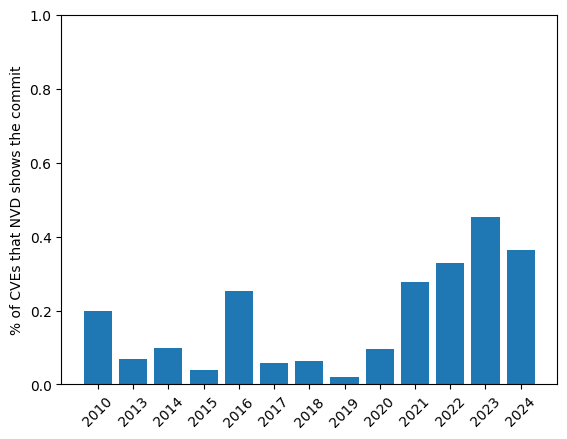}
    \caption{The proportion of available patches in NVD compared to GitHub Advisory database (average: 0.3). That is, for all the CVEs in GitHub Advisory with at least 1 patch link, we examine their corresponding NVD pages where the patch link is not missing.\label{fig:empirical_study} }
\end{figure}


In this section, we conduct an empirical study on the availability of patch links in NVD. 
To the best of our knowledge, we are not aware of any statistics on CVEs that guarantee to have a patch. 
Rather than reporting the absolute proportions of missing links, we find it more meaningful to compare the availability of links on one platform with another. 
The reason is that not all CVEs have a patch: for some CVEs, a patch may not be available yet. 
To make sure the studied CVEs have at least one patch, we compare the patch availability of the NVD database with the reviewed CVEs in the GitHub Advisory Database~\cite{githubAD}. The information of the reviewed CVEs (including the patch link) is verified by GitHub administrators, thus it guarantees the correctness of the metadata. 
All reviewed CVEs in the GitHub Advisory Database contain at least one patch link. More specifically, we investigate all the 2,000 reviewed CVEs under the Maven ecosystem~\cite{githubAD_maven}. For each CVE, we extract all the hyperlinks on the corresponding NVD webpage. Then we use regular expression to detect whether at least one hyperlink follows the format of a GitHub commit. We report this proportion (broken down by year) in Figure~\ref{fig:empirical_study}. 

\textbf{Results Analysis of Figure~\ref{fig:empirical_study}}. In Figure~\ref{fig:empirical_study}, the average proportion of the available commit links in NVD is 0.3. That is, the link is missing in 70\% CVEs. On the other hand, by observing the yearly trend, we can see that the proportion of NVD commits links has increased in the more recent years. For most of the years before 2020, the availability is lower than 10\%. These results show that compared with GitHub Advisory, it is more difficult to trace the CVEs in the NVD database, especially for earlier years. Such statistics confirm that the shortage of manpower has caused significant delays in NVD data maintenance. It is imperative to investigate the automated approach for retrieving the relevant patch of a CVE. 

\section{Problem Formulation}
\label{sec:formulation}

In this section, we formally define the problem of the explainable tracing of CVE patches in Open-Source software.
Given all the available information about a CVE, the problem studies how to identify the commit that fixes the vulnerability. After identifying the repository name, e.g., \texttt{apache/tomcat}, the problem is reduced to ranking all the commits from the repository. The query and candidates of the retrieval problem are explained below. 

\textbf{Query: CVE Information}. For the query, we can leverage all the available information regarding a CVE, including the CVE description, the repository name, and the version of the patch. An example of CVE description can be found in Figure~\ref{fig:cve_example}. The text description often mentions a natural language summary of the vulnerable code behavior, the affected software name, and the affected/patched software version, although not guaranteed to contain all fields~\cite{sun2024tracing}. The software name and version can help narrow down the candidates to the commits between the version ranges in the repository (We discuss this issue in Section~\ref{sec:data_collection:candidate_filter}). The software name and software version of a CVE can be automatically extracted from the CVE description using natural language processing techniques~\cite{fastxml,lightxml,yang2021few}. Meanwhile, most CVE databases~\cite{nvd,synk,githubAD} usually already provide the labeled software name and version. An example is the CPE information on an NVD page, e.g., "\emph{Up to (excluding) 8.5.7}". 

The text of a CVE description usually contains two parts of information that can be used for matching the relevant candidate. First, a high-level summary of the vulnerability e.g., the description of CVE-2017-6056~\cite{CVE-2017-6056} mentions "\emph{...It was discovered that a programming error...may result in denial of service via an infinite loop...}", which explains the vulnerability is \emph{Infinite Loop} or CWE-79~\cite{CWE-79}. Second, the descriptions often directly refer to the file name, function name, or variable name of the vulnerable code. For example, the description of CVE-2017-8085~\cite{CVE-2017-8085} mentions the PHP file containing the vulnerability: "\emph{In Exponent CMS before 2.4.1 Patch \#5, XSS in elFinder is possible in framework/modules/file/connector/elfinder.php}". 


\textbf{Candidate: Commit Information}. To represent each commit, we leverage the commit message text and the Diff code. Some commit messages of the patch contain a summary of the vulnerability, e.g., the commit message of the patch for CVE-2017-6056~\cite{CVE-2017-6056} is "\emph{...Fix potential infinite loop if pos == 0...}". The "\emph{infinite loop}" in this commit message overlaps with the CVE description above. On the other hand, the diff code contains information about the changed file, changed function, changed lines-of-code, and variables, which helps with the matching if the CVE description directly refers to them. 

\textbf{Explainable Information Retrieval}. Based on the above analysis, we propose to formulate the problem of patch tracing as an explainable information retrieval problem~\cite{singh2019exs,lyu2023listwise,anand2022explainable,singh2019exs}. Similar to explainable machine learning~\cite{ribeiro2016should}, explainable information retrieval highlights the important tokens in both the document and the query to help the searcher understand the ranking results of a retrieval model.

Given each query CVE $q$ and a list of candidate commits $d$, our task is to first retrieve the ranked list of commits, then highlight the tokens in $q$ and $d$ to help with the reviewing. If the highlighting algorithm is effective, it should help the CVE maintainer quickly distinguish the patch commit from the non-patched commits. 

\section{Construction of a Dataset for Security Vulnerability Patch Tracing}
\label{sec:data_collection}


In this section, we describe the data collection process. 

\subsection{Collecting CVEs and the Relevant Patch Commit Links}
\label{sec:data_collection:commits}

\textbf{Dataset Requirements and limitations of the current dataset.} To the best of our knowledge, no existing dataset can be directly used for our task of retrieving the relevant commit patch. The closest dataset is Sun et al.~\cite{sun2024tracing}, however, the SHA of the patch commit in their dataset is missing, and it is non-trivial to recover this SHA based on their dataset. 

\textbf{Data Sources and Collecting Process.} To collect more CVEs and more patch links, we use the combination of the following datasets: NVD~\cite{nvd}, BigVul~\cite{Fan2020BigVul}, patch\_db~\cite{wang2021PatchDB}, GitHub Security Advisory Database~\cite{githubAD}, and OSV Database~\cite{osv}. For BigVul and patch\_db, we use the labeled patch links. For NVD and GitHub advisory, 
we match GitHub commit links from the references as the patch commits for the current CVE. If multiple commits are extracted, the first one is selected. 
For OSV, we use its provided API to retrieve CVEs that are present in the databases but the GitHub commit links were not extracted in the previous databases.
We gather the union set of (CVE, patch commit) pairs from five different data sources. In the case of conflicting patch commits, priority is given to the BigVul and patch\_db datasets, as they provide labeled patch links. When neither BigVul nor patch\_db contain the relevant information, we use the information in NVD and GitHub advisory. 

While the patch can be from other websites, we focus on using the GitHub links, because it is easy to extract information using the GitHub API and related tools. 

Among the 9,435 CVEs collected from the 5 sources, we successfully identified patch commits for 7,242 CVEs out of the 9,435 CVEs.

\textbf{Removing Incorrect Commits}. 
After collecting the (CVE, patch) pairs, we remove incorrect cases. 
The incorrect reasons are: (1) repositories that no longer exist or are inaccessible, (2) repository restructuring that results in the patch commit being excluded from the main tree, and (3) commit hash inconsistencies where the provided commit hash in the vulnerability databases does not match the actual patch commit. 
We further exclude repositories that are too large, e.g., \texttt{torvalds/linux}. 
After this filtering process, our final dataset contains 3,573 CVEs. This filtering ensures that our dataset consists of verifiable and accessible patch commits, which is crucial for both training and evaluation purposes.

\subsection{Collecting the Candidate Commits}
\label{sec:data_collection:candidate_filter}

Since the number of candidate commits in a repository is large, we need to reduce the candidate size before retrieving the patch commit. We observe that most databases (e.g., NVD, GitHub Advisory, Snyk) provide a labeled patch version information. For example, for CVE-2017-8085~\cite{CVE-2017-8085}, the CPE information says the affected version is up to 2.4.0 (including), therefore the patch commit should locate between 2.4.0 and the next version. Technically, as long as this information is correct, we can automatically filter the candidate commits by version. However, in practice, we find that version information in NVD can suffer from various data quality issues, including imprecise version records (e.g., using broader release versions instead of specific patch versions), incorrect version records, misalignment between repository tags and version numbers, and completely missing version information. These data quality issues will greatly affect the coverage of the method, so we end up constructing a dataset by assuming we already know the patch commit and using the nearby commits as the candidate commits. We report the findings of both methods below.


\textbf{Method 1: Automatically Filtering Candidate Commits based on the NVD Version Tag.} There exist two major challenges in mapping the labeled tag (e.g., 2.4.0) to the range of tags for filtering (e.g., 2.4.0 to 2.4.1). First, if the labeled tag is the lower bound, how to find the upper bound and vice versa? To address this challenge, we use Python's \texttt{packaging} \cite{pythonPackaging} library with regular expressions to normalize and compare version strings, which allows us to identify the next/previous version based on the current version information. Second, if the labeled tag (e.g., 2.4.0) has a different name than the tag in the original repository (e.g., v2.4.0), how to resolve the name difference? For this challenge, we apply \texttt{difflib}'s\cite{pythonDifflib} \texttt{SequenceMatcher} algorithm. 


\textbf{Coverage Analysis of Method 1.} Out of all the 3,573 CVEs identified in Section~\ref{sec:data_collection:commits}, only 2,784 contain the version tags. We manually inspected these 2,784 CVEs and found that only 1,805 contain the accurate patch version information in NVD. For the 1,805 CVEs, our tag matching using \texttt{difflib} has a  92.9\% success rate, resulting in 1,688 CVEs. When the NVD tag is incorrect, we can increase the recall for covering the patch by increasing the candidate window to several versions before and after the NVD tag. We report this recall in Table~\ref{tab:distance_recall}. Results in Table~\ref{tab:distance_recall} indicate that while our fuzzy matching approach can effectively handle the version format differences, the actual coverage is significantly limited by version information quality in NVD. As a result, when preparing the candidate commits, we end up leveraging the true version tag. Although this information is not provided in the real application, we argue that our focus is to investigate the explainable tracing problem.


\begin{table}[h]
\centering
\caption{Statistics of distance range and recall \label{tab:distance_recall}}
\raisebox{0pt}[\dimexpr\height-0.5\baselineskip\relax]{
\begin{tabular}{p{1.2cm}p{1.2cm}p{1.2cm}p{1.5cm}} \hline
Range & Count & Total & Recall (\%) \\ \hline
0 & 1,677 & 2,784 & 60.24 \\
$\pm$1 & 1,875 & 2,784 & 67.35 \\
$\pm$2 & 2,028 & 2,784 & 72.84 \\
$\pm$3 & 2,083 & 2,784 & 74.82 \\
$\pm$5 & 2,136 & 2,784 & 76.72 \\
$\pm$10 & 2,263 & 2,784 & 81.29 \\
$\pm$50 & 2,614 & 2,784 & 93.89 \\ \hline
\end{tabular}
}
\end{table}


\textbf{Method 2: Collecting Candidate Commits based on the True Tag (the Method we Actually Used).} 
First, we find all tags that contain the patch commit. If we find multiple tags, we select the earliest one as the fixed version since it represents the first version which includes the patch commit. Next, we retrieve the version that comes right before the fixed version, which is the last vulnerable version.
Then, we collect all the commits between these two versions. These commits are our candidate pool of the potential patch commits because they represent all the changes between the last vulnerable version and the patched version. 

\textbf{Data Collection Results.} For each collected candidate commit, we extract commit messages and metadata, modified file paths and file contents, and detailed code changes showing additions and deletions. By utilizing version tags to limit our search range, we significantly reduced the number of commits that need to be examined from 70,722,387 total repository commits to 1,310,068 candidate commits, representing a 98.1\% reduction in the search space. 

We report the statistics of our dataset in Table~\ref{tab:stat}. Table~\ref{tab:stat} shows that the positive and negative ratio is highly skewed. The diff code is usually a lot longer than the commit message. 



    \begin{table}[h]
\caption{Statistics of the dataset \label{tab:stat}}
\centering
\begin{tabular}{cc} \hline
Avg tokens (cve desc) & 67.3 \\ 
Avg tokens (commit msg)& 43.9 \\ 
Avg tokens (diff code) & 1840.5\\
Median tokens (diff code) & 645\\
3-rd Quartile tokens (diff code) & 1846 \\\hline
\# commits & 114,523 \\
\# train & 40,827 \\
\# validation & 22,276 \\
\# test & 51,420 \\ \hline
Pos:neg & 1:45 \\ \hline
\end{tabular}
\caption*{\footnotesize For token count we use the tokenizer of CodeBERT~\cite{feng2020codebert}}
\end{table}

\section{Building an Explainable Patch Retrieval System}
\label{sec:methodology}

In this section, we describe the investigated framework for the explainable tracing of the patch commits. First, in Section~\ref{sec:training}, we describe how to train a multi-modal language model for the retrieval task (i.e., natural language and code). Second, in Section~\ref{sec:lime}, to identify the highlighted token, we first apply LIME~\cite{ribeiro2016should}, a highlighting token selection algorithm which is widely used for explainable machine learning~\cite{wu2023interpretable,halilovic2022explaining}, including explainable information retrieval~\cite{singh2019exs}. Third, we observe that some tokens highlighted by LIME are less informative, therefore we propose another highlighting method called TfIdf-Highlight (Section~\ref{sec:tfidf_highlight}). We describe each step as follows. 

\subsection{Training a Multi-Modal Language Model for Tracing Security Vulnerability Patches}
\label{sec:training}

\textbf{Discussion on Retrieval Model Selection}. Following existing work on similar tasks~\cite{sun2024tracing,chen2023identifying}, we train the retrieval model by fine-tuning language models, e.g., CodeBERT~\cite{feng2020codebert} and UnixCoder~\cite{guo2022unixcoder}\footnote{In this paper, we focus on investigating smaller models. However, our framework can be generally applied to large language models such as Llama models. }. Notice that an alternative approach is to use vector-based retrieval models~\cite{reimers2019sentence,karpukhin2020dense}. That is, first converting each of the CVE descriptions and commits into a vector, then training the retrieval model. We opt to not use the vector-based approach for the following reasons. First, compared with neural network, the vector-based models are harder to explain.
Second, after reducing the candidate commits size by tag filtering, the computational cost is largely reduced. Therefore, we do not have to resort to faster solutions using the vector-based models.

\begin{figure}[h]
 \centering
    \includegraphics[width=0.4\textwidth]{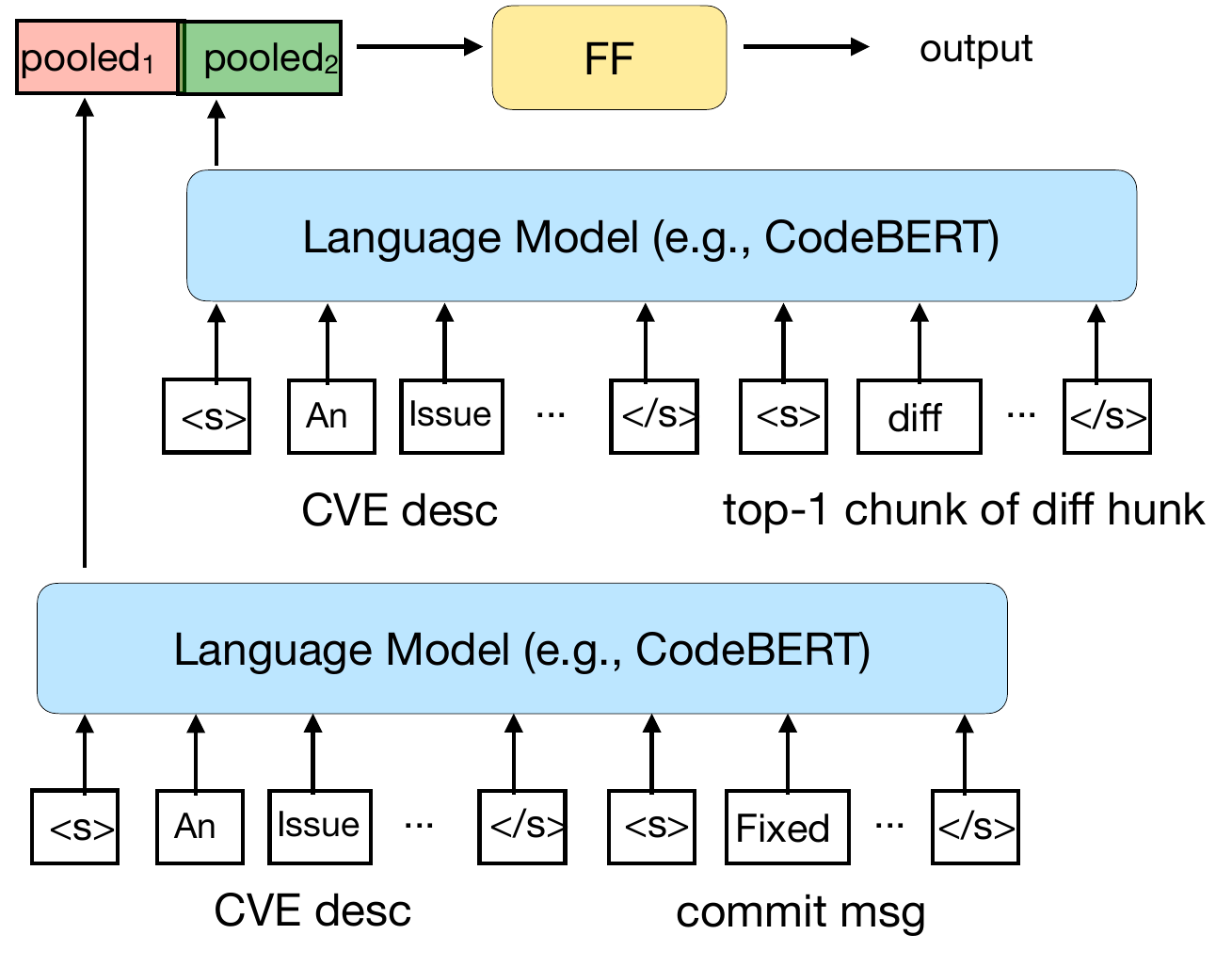}
    \caption{The retrieval model for the multi-modal training \label{fig:multimodal} }
\end{figure}

\textbf{Removing Whitespace}. We directly use the tokenizers of CodeBERT and UnixCoder. However, the raw tokenized outputs for code contain many whitespaces and a large number of tokens. To reduce the context length, we remove all the whitespaces (including line breaks and tabs). Although whitespaces are critical for defining indentations, removing them does not affect the token-level matching of the retrieval model. 

\textbf{Handling Long Contexts of Diff Code}. Since CodeBERT and UnixCoder both have a context length of 512 tokens, it is difficult to handle long contexts of diff code without truncation. To handle the long context challenge, we split each diff into chunks of 64 tokens. We use a shorter length of chunks to: (1) reduce the chance of truncation for longer CVE descriptions; and (2) make it easier for LIME to select the rationale tokens from the chunk. Then we rank all chunks based on their Tf-Idf cosine similarity (explained in Section~\ref{sec:tfidf_highlight}) with the CVE description. We only use the top-1 chunk in the model. 

\textbf{Multi-Modal Training Framework}. After the pre-processing, each commit is left with the commit message and one code chunk. We use the model in Figure~\ref{fig:multimodal}. For each pair of (CVE, commit), we first use the model to obtain the pooled embedding vector for each of the text and code, then feed the concatenate pooled vector to the classifier on top. Since the data distribution is imbalanced, the weight of positive:negative examples in the training data is set to 10:1. 

\textbf{Multi-Modal Training: Handling Modality Bias}. During the training of the model in Figure~\ref{fig:multimodal}, we identify modality bias~\cite{guo2023modality}. That is, when one modality (the code) is harder to train than another (the commit message), the model discovers this shortcut and only learns the easier modality. In particular, we find that when tested with the code + an empty commit message, the model's probability is nearly 0 (0.0269) for all positive cases. However, when tested with the commit message + an empty code, this problem does not happen. To force the model to learn from the code, we leverage data augmentation. For each positive example, we add 2 new examples: one with code only and one with commit message only. That is, if the commit is the patch, the model should recognize it even if only the code (commit message) is provided. We further add 10 times the single-modal negative examples to keep the positive:negative ratio consistent. After the data augmentation, the model's positive probability for code + an empty commit increases to 0.312. 

\subsection{Explaining Patch Ranking Results using LIME}
\label{sec:lime}

We first investigate using the LIME (locally interpretable model-agnostic explanation) algorithm~\cite{ribeiro2016should} to select the highlighted tokens. Given the trained model, for each test example \textbf{x} and a user-specified label (either 0 or 1), LIME can select $k$ tokens in the test example to explain the reason why the model predicts that label. The way LIME selects the $k$ tokens is by perturbing the test example into $n$ alternative samples \textbf{x}', and using the model to obtain $n$ probability scores $p$. Then LIME fits a ridge regression model (i.e., sparse linear regression) to $p$ (if label = 1) or 1-$p$ (if label = 0) by selecting $k$ features.   

\textbf{One-Label Explanation Problem}. Since our problem is a retrieval problem, i.e., for each CVE, rank all commits based on how likely they predict 1, we focus on always using LIME to explain 1 for all examples. 

\textbf{Commit/Code Only Highlighting}. Since our problem is a retrieval problem, we focus on using LIME to explain the commit message and diff code only. That is, for each example, when perturbing \textbf{x}$\rightarrow$ \textbf{x}', we do not perturb the tokens in the CVE description. We set this constraint so we could control the number of tokens perturbed in the commit message and diff code. For the same reason, the tokens in the commit message and diff code are perturbed separately. That is, for each test example, we run LIME twice, each selecting $k$ tokens from the commit message and diff code. 

\textbf{Dropping Non-informative Tokens}. To prevent LIME from selecting non-informative tokens, we further restrict that the selected token must contain at least one letter or number. 

\subsection{TfIdf-Highlight}
\label{sec:tfidf_highlight}

Although LIME can select $k$ highlighted tokens, we observe that LIME sometimes selects non-informative tokens, e.g., meaningless words such as \emph{are}, \emph{its}, and \emph{or}. It is a non-trivial problem to remove all the non-informative words (e.g., using rules or a stopword set). Therefore, we propose another approach for selecting the highlighted tokens. 

Our approach, TfIdf-Highlight, is simply inspired by the Tf-Idf retrieval model. Given the CVE $q$, the commit message $msg$, and the diff code $diff$, for any word $w \in q$, we use the following TfIdf score to rank the words in $msg$ and $diff$, where the top $k$ words are selected as the rationale tokens:


\begin{align*}
    \text{logtf}(w, d) &= 0.5 \log_2\{\text{tf}(w, d_{msg})\} + 0.5 \log_2\{\text{tf}(w, d_{diff})\} \\[0.5em]
    \text{idf}_{\text{CVE}}(w, q) &= \max\{\log_2(\frac{N_q}{\text{df}_{\text{CVE}}(w, q)}) - \alpha, 0\} \\[0.5em]
    \text{idf}_{\text{all}}(w) &= \max\{\log_2(\frac{N}{\text{df}(w)}) - \beta, 0\} \\[0.5em]
    \text{TfIdf}(w, d_{msg}) &= \text{logtf}(w, d) \cdot \text{idf}_{\text{CVE}}(w, q) \cdot \text{idf}_{\text{all}}(w) \\
    &\quad \text{if } w \in d_{msg} \cap q \\[0.5em]
    \text{TfIdf}(w, d_{diff}) &= \text{tf}(w, d) \cdot \text{idf}_{\text{CVE}}(w, q) \cdot \text{idf}_{\text{all}}(w) \\
    &\quad \text{if } w \in d_{diff} \cap q
\end{align*}

Here, $\text{tf}(w, d_{msg})$ is the term frequency of word $w$ in the commit message, $\text{tf}(w, d_{diff})$ is the term frequency of word $w$ in the diff code, $\text{df}_{\text{CVE}}(w, q)$ is the number of commits under the CVE $q$ which contain the word $w$, $N_q$ is the total number of comments under the CVE $q$, and $\text{df}(w)$ is the number of commits in the entire dataset that contain the word $w$. Different from the standard Tf-Idf approach, here we introduce $\text{idf}_{\text{CVE}}(w, q)$ to penalize words that are less informative within the CVE. For example, words describing the repository name (e.g., "\emph{rave}" in Figure~\ref{fig:cve_example}) are non-informative within the repository but are informative outside the repository. Here we set $\alpha = 1$ and $\beta = 0.01$: if more than half of commits within the CVE contain the word, the word is not informative.

\textbf{Tokenization using Snake/Camel Case}. To match more tokens, we leverage regular expression to split all fields using the Snake/Camel case. We keep both the tokens before the split and after the split. For example, for the word \emph{assertEquals}, 3 tokens are added: \emph{assert}, \emph{Equals}, and \emph{assertEquals}. 

After obtaining the highlighted tokens for each commit message and diff code, we map the highlighting result to the token space of the language model (e.g., CodeBERT and UnixCoder) using character indices. For each token by CodeBERT, the TfIdf-Highlight score is equal to the max of all overlapped original tokens. 

\section{Evaluation}

In this section, we study the effectiveness of the proposed explainable tracing framework (Section~\ref{sec:methodology}). Section~\ref{sec:eval_retrieval} evaluates the trained model's retrieval performance. Section~\ref{sec:eval_explainability} compares the rationale selection methods in Section~\ref{sec:eval_explainability}. Finally, Section~\ref{sec:eval_human} describes a blind human labeling experiment for comparing the rationale selection methods. 

The research questions we investigate are the following:

\begin{itemize}
\item \textbf{RQ1}: How effective is the trained model for retrieving the patch commit?
\item \textbf{RQ2}: What are the explanation performances of LIME and TfIdf-Highlight?
\item \textbf{RQ3}: How do explanations affect human annotators's decision-making process?
\end{itemize}

\subsection{Evaluating the Retrieval Performance}
\label{sec:eval_retrieval}

\textbf{Experiment Settings}. We run all experiments on a cloud-based computing server (Ubuntu 22.04), with an RTX 4090 GPU (24GB). 

\textbf{Model Selection}. In this paper, we study two language models for retrieving the patch commit: CodeBERT-base~\cite{feng2020codebert} and UnixCoder-base~\cite{guo2022unixcoder}. We choose CodeBERT because it has the best retrieval performance for the file-level patch-tracing task in Sun et al.~\cite{sun2024tracing}. Both CodeBERT and UnixCoder are frequently used for vulnerability-related tasks~\cite{wang2024m2cvd,zhou2021finding,zhou2024large} including the retrieval of code~\cite{compvpd}. While existing work shows that smaller language models (e.g., RNN, Bi-LSTM) are easier to explain, we skip these models since they usually have worse performance. We leave exploring large language models (e.g., Llama) for future work. 

\textbf{Hyperparameter Settings}. For the model training, we use a learning rate = 2e-5, batch size 
= 8, epoch = 5, and max\_length = 256. Since the code is split into chunks of 64 tokens and we only use the top-1 chunk (Section~\ref{sec:training}), only 4\% text and 0.7\% code are truncated into 256 tokens. This design is for keeping most tokens within the search space for highlighting. 

\begin{table}[h]
\caption{Comparing the retrieval performance of different models \label{tab:ranking_perf}}
\raisebox{0pt}[\dimexpr\height-0.5\baselineskip\relax]{
\begin{tabular}{p{1cm}p{1cm}p{0.8cm}p{0.8cm}p{0.8cm}p{0.8cm}p{0.8cm}} \hline
\multicolumn{2}{c}{\multirow{2}{*}{Metric}} &  Tf-Idf &\multicolumn{2}{c}{CodeBERT} & \multicolumn{2}{c}{UnixCoder}  \\ \cline{3-7}
& & Test & Valid & Test & Valid & Test \\ \hline
\multirow{4}{*}{Classify} & Accuracy & - & 0.979 & 0.984 & 0.976 & 0.984\\ 
& Prec & - & 0.590 & \textbf{0.537} & 0.527 & 0.526\\ 
& Recall & - & 0.651 & 0.657 & 0.646 & \textbf{0.674}\\ 
& F1 & - & 0.619 & 0.591 & 0.580 & 0.591\\ 
& Auc & 0.877 & 0.932 & 0.964& 0.956 & \textbf{0.968}\\ \hline

\multirow{3}{*}{Rank} & Prec@1 & 0.507 & 0.652& 0.657& 0.636 & \textbf{0.658}\\ 
& Recall@1 & 0.507 & 0.652 & 0.657& 0.636 & \textbf{0.658}\\
& Recall@2 & 0.636 & 0.717 & 0.756& 0.725 & \textbf{0.760}\\
& Recall@5 & 0.754 & 0.780 & 0.825& 0.787 & \textbf{0.833}\\
& MAP & 0.689 & 0.824 & 0.817& 0.821 & \textbf{0.827}\\ \hline
\end{tabular}
}
\begin{minipage}{\linewidth}
        \raggedleft
\caption*{\footnotesize The pos:neg ratio is 1:59 in test and 1:39 in valid. }
\end{minipage}
\end{table}

\textbf{Baseline}. We compare CodeBERT and UnixCoder with Tf-Idf~\cite{tfidf}. For the tokenization of Tf-Idf, we leverage the same regular expression with Camel/snake case splitting as Section~\ref{sec:tfidf_highlight}. The weight for the commit message in Tf-Idf is 0.8 and for the diff code is 0.2. 

\textbf{Results Analysis}. In Table~\ref{tab:ranking_perf}, we compare the ranking performance of different models. Overall, we find that UnixCoder and CodeBERT both outperform Tf-Idf. UnixCoder and CodeBERT have similar scores for classification and ranking. Notice that unlike the blackbox language models, the decision of Tf-Idf can be easily explained using the Tf and Idf scores. Since the explainable Tf-Idf model has a worse performance than the non-explainable CodeBERT and UnixCoder models, the goal of our work is meaningful. That is, we try to improve the explainability of the model without compromising on the ranking performance.

\textbf{Summary of Findings for RQ1}. We find that CodeBERT and UnixCoder achieve similar retrieval scores for patch retrieval.

\begin{table*}[t]
\centering
\caption{Comparing the faithfulness of TfIdf-Highlight vs LIME~\cite{ribeiro2016should} by controlling the \#highlighted tokens \label{tab:highlight_perf}}
\begin{tabular}{p{1.2cm}p{1.2cm}p{0.8cm}p{0.9cm}p{0.9cm}p{0.9cm}p{0.9cm}p{0.9cm}p{0.9cm}p{0.9cm}p{0.9cm}p{0.9cm}} \\ \hline
\multirow{2}{*}{Model} &\multirow{2}{*}{Component} & \multirow{2}{*}{Fold}& $\overline{\#tokens}$(pos) & \multicolumn{4}{c}{Sufficiency$\downarrow$ } & \multicolumn{4}{c}{Comprehensiveness $\uparrow$}  \\ 
 & & & & LIME & TfIdf-Highlight & p-val & t & LIME & TfIdf-Highlight & p-val & t \\ \hline
\multirow{4}{*}{CodeBERT}& \multirow{2}{*}{Text only} & Valid & 3.16 & 0.398  & \textbf{0.286} & 4e-9 & -5.9 & 0.089 & \textbf{0.116} & 3e-3 & 2.89\\ 
& & Test & 3.22 &0.351 & \textbf{0.305} & 2e-3& -3.07& 0.125 & 0.139 & 0.115 & 1.57\\ \cline{2-12}
& \multirow{2}{*}{Code only} & Valid & 4.15 & 0.494 & \textbf{0.403} &3e-6& -4.68 & 0.057 & \textbf{0.081} & 0.016 & 2.4  \\ 
& & Test & 4.14 & 0.484 & \textbf{0.413} & 2e-6& -4.7 & 0.061& 0.075 &0.05 & 1.94 \\ \hline \hline

\multirow{4}{*}{UnixCoder} & \multirow{2}{*}{Text only} & Valid  &2.88 & 0.285 & \textbf{0.232} & 9e-4 & -3.3 & 0.112 & 0.125 & 0.238 & 1.17  \\ 
& & Test  & 3.06& 0.325 & \textbf{0.275}  &3e-4 &-3.61 & 0.136 & 0.140 & 0.722 & 0.355 \\ \cline{2-12}
& \multirow{2}{*}{Code only} & Valid  &3.70 & 0.432 & \textbf{0.344} & 5e-8& -4.57& 0.088 & 0.088& 1.0 & 0\\ 
& & Test & 3.79 & 0.424 & \textbf{0.366} & 3e-4& -3.5& 0.083 &0.083 & 0.884 & 0.145\\ \hline
\end{tabular}
\end{table*}

\begin{figure*}[h]
    \centering
    \begin{subfigure}{0.24\textwidth}
        \centering
        \includegraphics[width=\linewidth]{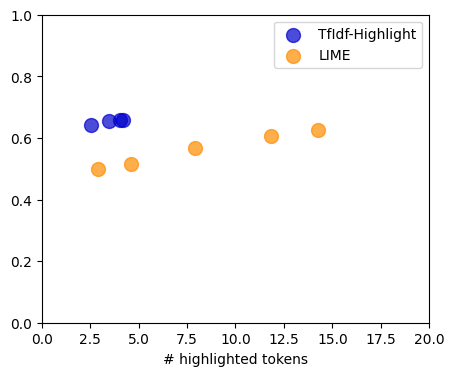}
       \caption{Pos prob w Highlight Text$\uparrow$}
    \end{subfigure}
    \begin{subfigure}{0.24\textwidth}
        \centering
        \includegraphics[width=\linewidth]{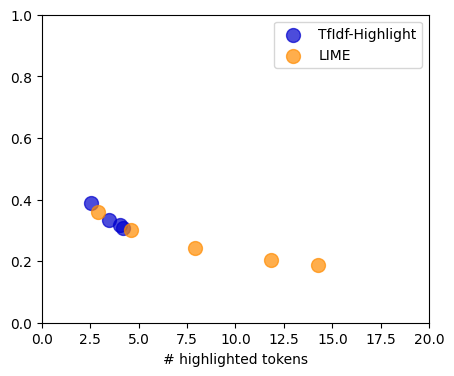}
       \caption{Pos prob wo Highlight Text$\downarrow$}
    \end{subfigure}
    \begin{subfigure}{0.24\textwidth}
        \centering
        \includegraphics[width=\linewidth]{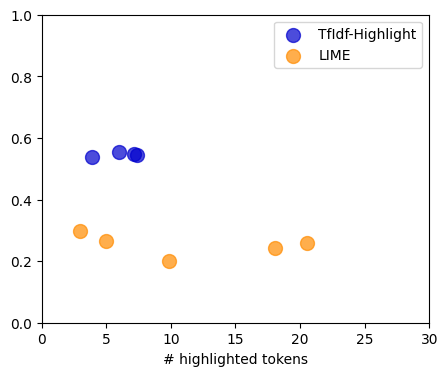}
       \caption{Pos prob w Highlight Code$\uparrow$}
    \end{subfigure}
    \begin{subfigure}{0.24\textwidth}
        \centering
        \includegraphics[width=\linewidth]{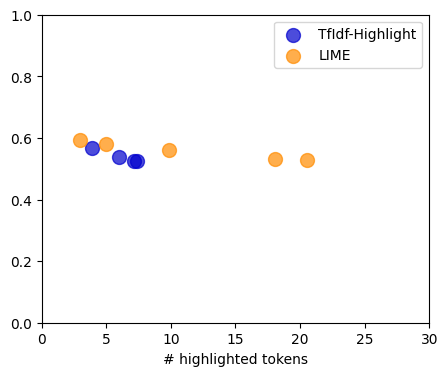}
      \caption{Pos prob wo Highlight Code$\downarrow$}
    \end{subfigure}
   \caption{The trend of explainability score vs. the number of highlighted tokens. Model: CodeBERT, Fold: valid\label{fig:explainability_vs_num} }
\end{figure*}

\subsection{Evaluating the Performance of Rationale Selection}
\label{sec:eval_explainability}

In this section, we evaluate the explanation performance of TfIdf-Highlight by comparing it with LIME~\cite{ribeiro2016should}. 

\textbf{Defining Faithfulness in Ranking}. Existing evaluation metrics for explainable machine learning often require comparison with ground truth tokens~\cite{deyoung2019eraser,mathew2021hatexplain}. However, since our dataset is large and manually labeling long code data is challenging (Table~\ref{tab:stat} shows the average length of the diff code is 1,840 tokens), we have not collected the ground truth labeling and will leave it for future work. One evaluation metric that does not require ground truth labeling is called \emph{faithfulness}, which captures the extent to which the highlighted tokens result in the same model decision as the full input~\cite{jacovi2020towards,mathew2021hatexplain}. For example, in Matthew et al.~\cite{mathew2021hatexplain}, faithfulness is defined as the distance between $proba[predicted\_label]$ and $proba'[predicted\_label]$, where $proba$ is the predicted probability using the original input, and $proba'$ is the predicted probability using only the highlighted tokens as the input (sufficiency), or excluding the highlighted tokens (comprehensiveness). However, their problem is a multi-class classification problem, which is different from our ranking problem. We define the faithfulness of ranking below:

\begin{eqnarray*}
faithfulness &= \frac{1}{n} \sum_{cve} |Prec@1(cve, proba) \\
&- Prec@1(cve, proba')|
\end{eqnarray*}

Here $Prec@1(cve, proba)$ is the precision@1 for each query $cve$ using $proba$ as the probabilities. This $Prec@1$ can be replaced by other ranking metrics. Similar to Matthew et al.~\cite{mathew2021hatexplain}, we compare the ranking using the highlighted tokens only probabilities $proba'$ (or excluding the highlighted tokens) with the original probabilities $proba$. Since we only explain either the commit message or the diff code only, but not both, $proba'$ is obtained by keeping the code the same as the full input, while using only the highlighted tokens for the commit message, etc. 

\textbf{Controlling the Number of Highlighted Tokens}. When comparing the faithfulness scores between LIME and TfIdf-Highlight, we need to keep the number of highlighted tokens the same for each example. This is because the higher the number of highlighted tokens, the closer to the original input, therefore the model will have a lower sufficiency score and a higher comprehensiveness score. While it is easier to control the number of highlighted tokens in LIME, TfIdf-Highlight will cut the highlighted tokens based on the Tf-Idf score, so the number of highlighted tokens is different for each example. To control the number of highlighted tokens, we first cut the highlighted tokens in Tf-Idf at $k$ (therefore Tf-Idf may highlight 0-$k$ tokens), then use LIME to highlight the same number of tokens as Tf-Idf. This setting guarantees a fair comparison between TfIdf-Highlight and LIME.

\textbf{Result Analysis}. We report the faithfulness scores in Table~\ref{tab:highlight_perf}. For each setting, we report the average number of highlighted tokens in the positive examples, the average sufficiency and comprehensiveness scores, and the statistical significance test scores. From Table~\ref{tab:highlight_perf} we can see that the sufficiency scores of LIME are worse than TfIdf-Highlight in almost all settings. The sufficiency metric measures the extent to which the highlighted tokens reflect the original model's decision-making process. While the LIME tokens are identified based on the model, TfIdf-Highlight does not need to know any information about the model, thus this result is surprising. 

By observing the values in Table~\ref{tab:highlight_perf}, we find that the text-only mode usually has a lower sufficiency score and a higher comprehensiveness score. This result meets our expectations because the commit message is usually more informative than the code for the retrieval. 

\textbf{Explainability Score vs. \#Highlighted Tokens}. To understand the trend of TfIdf-Highlight vs LIME under different numbers of highlighted tokens, we further show the plot in Figure~\ref{fig:explainability_vs_num}. The y-axis of Figure~\ref{fig:explainability_vs_num} is the average positive probabilities in $proba'$ (i.e., the probability with/without only the highlighted tokens) of all the positive examples (i.e., patch). From Figure~\ref{fig:explainability_vs_num} we can see that even with a few times as many highlighted tokens, it is difficult for LIME to achieve the same positive probabilities as TfIdf-Highlight. 

\textbf{Summary of Findings for RQ2}. We find that TfIdf-Highlight significantly outperforms BERT in the sufficiency score for all settings and slightly outperforms BERT in the comprehensiveness score.

\subsection{Human Labeling of the Rationale Selection}
\label{sec:eval_human}



While the explainability scores reflect a highlight's faithfulness to a model's decision-making, it does not reveal how helpful the tokens are for human reviewers to make a decision. Therefore, we conduct a user study experiment to examine the helpfulness of the highlights. 

\begin{figure}[h]
 \centering
    \includegraphics[width=0.4\textwidth]{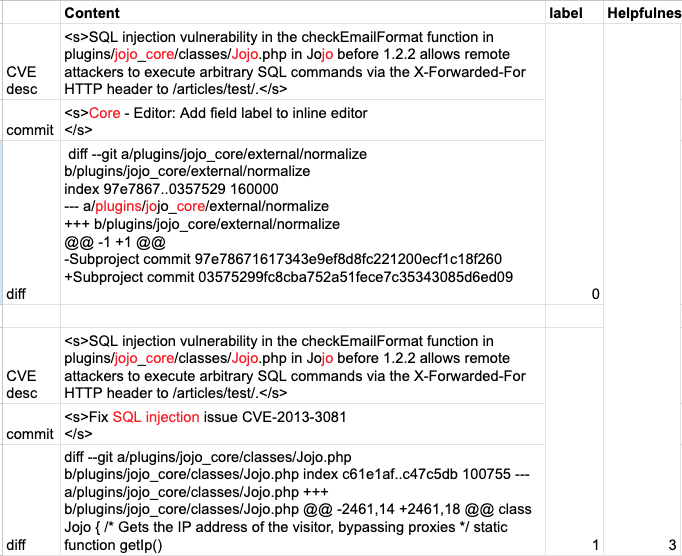}
    \caption{An example of the human labeling interface \label{fig:human_interface} }
\end{figure}



\textbf{Annotator Information}. Three authors separately provide the labels. Annotator 1 is a CS PhD student with knowledge on security vulnerabilities and frequent CWEs. Annotator 2 and Annotator 3 are CS undergraduate students with knowledge on security vulnerabilities.  


\textbf{First Blind Experiment (Easy)}. For the first experiment, we ask 2 annotators to guess the patch from 100 groups of (CVE, commit) pairs. All examples are highlighted, and an example of the interface of labeling is shown in Figure~\ref{fig:human_interface} (implemented using Google App Script\cite{google_app_scripts}. For each group of CVEs, we display the original commit message and the diff code, while highlighting the tokens selected by the highlighting algorithm. 

Each group contains 1 CVE and 2 commits. One commit is the patch and the other commit is the top-2 ranked commit using TF-IDF. The choice of top-2 is for increasing the challenge of the labeling task. The 2 commits in each group are randomly shuffled. For each group, we randomly assign one highlighting method to annotator 1 and assign the other method to annotator 2. The annotators are not told which highlighting method is used in each group. The annotators are asked to provide 2 labels: choose the patch from the 2 commits, and rate the helpfulness from 1-3 Likert scale (1: not helpful at all; 2: the highlighted token can explain the match, but incomplete; 3. the highlighted tokens are perfect for explaining the match). 
The results in Table \ref{tab:highlight_perf} indicate that TfIdf-highlight receives higher average helpfulness ratings than LIME. However, the accuracies of labeling with the two highlighting methods are similar. 

%


\textbf{Second Blind Experiment (Hard)}. The second experiment is similar to the first except the following differences: (1) adding no highlighting as the third setting (as a result we add a third annotator); (2)  if at least one commit is for changing .txt, .md, etc., we remove the entire group. This step is for eliminating cases that are too simple to label; (3) increasing the group size of commits from 2 to 4 to increase the challenge; (4) the number of groups is reduced to 50. (5) including two rounds: in the first round, the commit message is hidden, annotators need to select by reading the code only; in the second round, all information is revealed (including the first round label), they need to label whether to change the first round label. Same as the first experiment, the highlighted and non-highlighted groups are randomly shuffled and not revealed. 

The result of the second experiment can be found in Table~\ref{tab:human_labeling_results_2}. We can make the following observations: first, the overall accuracy of with highlighting is actually lower than without highlighting. Second, by comparing TfIdf-Highlight with LIME, we find that when labeling with code only, TfIdf-Highlight has aa worse accuracy than LIME, but after revealing the commit message, their accuracies are the same. This result shows that commits are critical for improving the accuracies. By comparing the three annotators, we find that the PhD annotator 1 who has knowledge on CWEs has a better accuracy than the two undergraduate annotators. 

\textbf{Case Study of Wrong Guesses}. We inspect the reasons why the accuracy of without highlighting is higher than with highlighting. By interviewing the annotators, we find that the mental model they use for making decisions has 2 steps: first, they use the keywords in the CVE description and look for the exact match information (e.g., file names, variable names) in the commits to handle the easy cases; second, if the first step feels hard, they look for the changed lines of code for a deeper understanding of the vulnerability, and see if it matches the vulnerability statement in the CVE description. While highlighting (especially TfIdf-Highlight) shows the token-level matching, it cannot reflect a deeper understanding of the vulnerability. For example, CVE-2020-5295 is "\emph{$\cdots$, an attacker can exploit this vulnerability to read local files of an October CMS server. $\cdots$}", and the patch commit includes words related to file operation such as \emph{file}, \emph{directory}, and \emph{path}. Tf-Idf does not highlight the word \emph{file} because it is a frequent word in the entire dataset, it also misses the other words due to the token mismatch. In general, a negative commit may have more matched tokens than the patch commits, thus highlighting can indeed be a double-sided sword in distracting the annotator from looking for the correct rationale tokens. 



\textbf{Summary of Findings for RQ3}. We find that the reasoning process for selecting the patch requires a deeper understanding of the semantic relatedness between the CVE description and the commit. Highlighting based on only token-level matching using Tf-Idf may not be sufficient to support the patch tracing task.

\begin{table}
\centering
\caption{Results of the first user experiment (easy) \label{tab:human_labeling results}}
\begin{tabular}{p{0.7cm}p{0.8cm}p{2cm}p{1cm}p{1cm}} \\ \hline
Labeler & Remove Changelog & Method & Accuracy & helpfulness\\ \hline
\multirow{3}{*}{1} & \multirow{3}{*}{True} & TfIdf-Highlight & 0.92 & \textbf{2.30}  \\ 
          &   & Lime & \textbf{0.96} & 2.05 \\ 
          &   & Total &  0.94 & 2.20 \\ \hline
          
\multirow{3}{*}{2} & \multirow{3}{*}{True} & TfIdf-Highlight & \textbf{0.93} & \textbf{2.46}  \\ 
          &   & Lime & 0.88 & 2.11  \\
          &   & Total & 0.91 & 2.31 \\\hline
\multirow{3}{*}{overall} & \multirow{3}{*}{True} & TfIdf-Highlight & 0.92 &\textbf{2.40} \\ 
          &   & Lime & \textbf{0.93} & 2.08  \\ 
          &   & Total & 0.92 & 2.24 \\\hline
\end{tabular}
\end{table}

\begin{table}
\centering
\caption{Results of the second user study experiment (hard)\label{tab:human_labeling_results_2}}
\begin{tabular}{p{1cm}p{1.5cm}p{0.8cm}p{0.8cm}p{0.8cm}p{0.8cm}} \\ \hline

\multirow{2}{*}{Labeler} & \multirow{2}{*}{CommitMsg} & \multicolumn{3}{c}{Method}   & \multirow{2}{*}{Total} \\
          &               & TfIdf-Highlight  & Lime & None &       \\\hline
\multirow{2}{*}{1}         & True          & \textbf{1.0}    & 0.72 & 0.82 & 0.84  \\
          & False         & 0.8    & 0.77 & \textbf{0.88} & 0.82  \\\hline
\multirow{2}{*}{2}         & True          & 0.43   & \textbf{0.91} & \textbf{0.91} & 0.76  \\
          & False         & 0.56   & \textbf{1.0}  & 0.87 & 0.8   \\\hline
\multirow{2}{*}{3}         & True          & \textbf{0.84}   & 0.71 & 0.8  & 0.78  \\
          & False         & 0.63   & 0.57 & \textbf{0.7}  & 0.62  \\\hline
\multirow{2}{*}{Overall}   & True          & 0.76   & 0.76 & \textbf{0.86} & 0.79  \\
          & False         & 0.66   & 0.74 & \textbf{0.84} & 0.75 \\ \hline
          
\end{tabular}
\end{table}








\section{Related Work}


In this section, we summarize the related work. 

\textbf{Security Vulnerability Database Maintenance}. To assist security personnel in detecting, maintaining, and tracing a security vulnerability, centralized databases are constructed to keep track of security vulnerabilities
The two major databases for security vulnerabilities are NVD~\cite{nvd} and MITRE~\cite{mitre}, which maintain synchronized information about common vulnerabilities and exposures (CVE). One critical task for security vulnerability databases is to keep track of critical metadata information to help users be aware of the update, e.g., the version number, the software name~\cite{fastxml,yang2021few,lightxml}, and the fine-grained affected package name (e.g., Java)~\cite{chen2023vullibgen,chen2023identifying}. For example, by identifying the affected package name, users who use the package can be alerted when there is an update of the CVE. 

In recent years, due to the tremendous growth of CVEs, the manual maintenance of vulnerability databases has become more and more challenging~\cite{nvd_delay,nvd_delay2,nvd_delay3}. To this end, existing work builds automated systems to accelerate the maintenance tasks. For example, existing work uses named-entity recognition (NER) to extract the software name and version information~\cite{yang2021few,dong2022survey,anwar2021cleaning,kuehn2021ovana}. Other work also models the problem as extreme multi-class classification (XML) where each software name is selected from a pre-defined list of packages~\cite{lightxml,fastxml,chronos}. However, the XML approach can only handle thousands of package names. To identify fine-grained software packages from an even larger number of candidates (e.g., Java packages), Chen et al.~\cite{chen2023identifying} train a retrieval model using the BERT model, which can identify the affected package with high accuracy when there are tens of thousands of candidates. Chen et al.~\cite{chen2023vullibgen} further leverage large language models to directly generate the affected package names. 


\textbf{Explainable Machine Learning}. In recent years, deep neural network models have been widely applied to applications in all industries. However, one major challenge with deep learning models is the lack of transparency in the output~\cite{ribeiro2016should,guo2018lemna}. When deploying an automated machine learning system, it is critical to know that the model is making the right decision in order to build trust between humans and the model. Matthew et al.~\cite{mathew2021hatexplain} uses explainable machine learning on hate speech detection tasks. They find that smaller models such as RNN and Bi-LSTM are easier to explain than larger models such as BERT. They use regularization to force the trained model's attention to focus on the labeled tokens. They find the model's explainability scores improve after training with regularization. Guo et al.~\cite{guo2018lemna} develop an explainable machine-learning learning model algorithm and apply it on security applications including binary code analysis and classifying PDF malware. Explainable machine learning has also been used for anomaly detection to capture an attacker's strategy and help the user make better decisions~\cite{li2023survey}. 

Ribeiro et al.~\cite{ribeiro2016should} find that the highlighted tokens can help non-expert human subjects make better decisions on classification tasks. Alufaisan et al. finds that users tend to follow the AI prediction when it is available~\cite{alufaisan2021does}. Previous work finds that humans struggle with interpreting the explanation when the highlighted areas do not align with the human's intuition~\cite{molnar2020general,knab2024dseg}.  Besides highlighting the tokens (i.e., hard explanation), existing work also defines explanations as a set of soft rules~\cite{chen2020denas}. 


\section{Discussions and Future Work}

\textbf{Combining Explanations with Domain Knowledge}. The user study results in Section~\ref{sec:eval_human} show that highlighting may negatively affect the accuracy of labeling. To support users in making the right decision, the highlight has to capture the semantic relatedness between the CVE description and the commit. To this end, one future work direction is to leverage the security domain knowledge using knowledgebase information such as the common weakness enumerations (CWE). 

\textbf{Using Large Language Models to Enhance Explanations}. Another format of explanation is to use natural language to explain the relation between the CVE description and the commit. We can leverage large language models to generate rationales for the match. For example, if the annotator does not understand why the highlighted tokens reflect a vulnerability, we can explain the general rules to bridge the knowledge gap of the annotator. 


\section{Threats to Validity}

The statistics on NVD patch availability (Section~\ref{sec:empirical_study}) are based on the comparison with 2000 CVEs on Maven in the GitHub Advisory database. When using other CVEs, this result may change. 

Our study has excluded the CVEs and commits in large repositories such as \texttt{torvalds/linux} (Section~\ref{sec:data_collection}) because their number of commits between versions greatly exceeds our processing capacity. We are capable of collecting enough data to validate our idea while keeping the computational demands under control using this selective approach. 

Our user study focuses on asking three annotators to label 50-100 groups of commits. Since the number of labeled CVEs is relatively small, the conclusion may not stay true when the human subjects are a different group of people. 

\section{Conclusion}

In this work, we conduct the first study on using explainable retrieval for tracing the vulnerable patch of a security vulnerability. First, we construct a dataset for the automated tracing of a security vulnerability patch. Second, we train a multi-modal retrieval model for the commit message and diff code given the CVE description. Third, we apply LIME, a widely applied explainable machine learning algorithm to the patch tracing problem. Fourth, we propose a new explanation algorithm called TfIdf-Highlight, which leverages the Tf-Idf information to find the most informative words in the repository and the entire dataset. 

Our evaluation shows that TfIdf-Highlight consistently outperforms LIME on the faithfulness score. That is, compared with LIME, the tokens selected by TfIdf-Highlight truly reflect the decision-making process of the trained model. By asking three annotators to blindly label the patch, we find that TfIdf has a similar labeling accuracy to LIME and a higher helpfulness score. Nevertheless, the accuracy of highlighting is lower than non-highlighting, since the highlighted tokens only reflect the token-level matching. To help with the decision-making, the highlighting needs to go beyond token matching to reflect a deeper understanding of the security domain knowledge.

\clearpage

\bibliographystyle{IEEEtran}

\bibliography{msr}

\end{document}